\documentstyle[preprint,aps]{revtex}
\begin{document}
\title{Proton drip-line nuclei in 
relativistic mean-field theory}
\author{G. A. Lalazissis and S. Raman}
\address{Physics Division, 
Oak Ridge National Laboratory, Oak Ridge, Tennessee 37831}
\date{\today}
\maketitle
\begin{abstract}
The position of the two-proton drip line has been calculated 
for even-even  nuclei with $10 \leq  Z \leq 82$ in the framework  
of the relativistic mean-field (RMF) theory. The current model uses
the NL3 effective interaction in the mean-field Lagrangian 
and describes pairing correlations in the 
Bardeen-Cooper-Schrieffer (BCS) formalism. 
The predictions of the RMF theory are compared with those of the 
Hartree-Fock+BCS approach (with effective force Skyrme SIII)  
and the finite-range droplet model 
(FRDM) and with the available experimental information.
\end{abstract}
\vspace{1 cm}
%
\section {Introduction}
Experimental and theoretical studies of exotic nuclei with
extreme isospin values are active areas of current research 
in nuclear physics. The advent of radioactive beams and the creation of 
several facilities to produce them have provided the opportunities to study
the structure and properties of very short-lived nuclei with extreme 
neutron-to-proton $(N/Z)$ ratios 
\cite{Rkl.92,MS.93,Gei.95,Han.95,Tan.95,Ver.96,Proc.96}. 

On the neutron-rich side, exotic phenomena include (i) the weak binding of 
the outermost neutrons, (ii) pronounced effects of the coupling between 
bound states and the particle continuum, and (iii) regions of neutron halos
with very diffuse neutron densities and major modifications
in the shell structures. The situation is different on the proton-rich side
of the stability valley. Here, nuclei are stabilized by the 
Coulomb barrier, which tends to localize the proton density in the nuclear 
interior, thereby preventing the 
formation of nuclei with large spatial extensions.

The opportunities provided by the radioactive beam facilities make 
the study of the structure and properties of nuclei close to the proton 
drip line a very interesting topic from both experimental and
theoretical points of view. Experimentally, possibilities for studying 
new decay modes such as diproton emission have opened up. 
Theoretical studies allow further tests of the various models. 
Of special interest is the region of $sd$--$fp$-shell proton-rich 
nuclei \cite{Orm.96,Orm.97,NaZ.96,VLR.98} where two-proton  
ground-state radioactivity \cite{Det.90,Bor.92,Bla.94,Bla.96} is expected 
to occur. In particular, the region around $^{48}$Ni is expected to 
contain nuclei which are two-proton emitters. 

In certain cases the proton drip line has been reached or
even crossed experimentally. Systematic theoretical studies predicting the
positions of the proton drip line are therefore important and 
timely \cite{Woods.97}.
In this work, the relativistic mean-field (RMF) theory is used 
to study the ground-state 
properties of very proton-rich, even-even nuclei with  
$10 \leq  Z \leq 82$ and to predict the location 
of the two-proton drip line. 

The RMF theory \cite{SW.86,Rei.89,Ser.92,Ring.96}
has proven to be a powerful tool to describe and predict
the properties of nuclei. This theory provides an elegant and economical
framework, in which properties of nuclear matter and finite nuclei,
as well as the dynamics of heavy-ion collisions, can be
calculated (for a recent review, see Ref.~\cite{Ring.96}).   
Compared to conventional nonrelativistic approaches,
relativistic models explicitly include mesonic degrees of
freedom and describe the nucleons as Dirac particles. Moreover,
the spin-orbit interaction arises naturally from the
Dirac-Lorenz structure of the effective Lagrangian.

In this work, 
the calculations are performed in the axially-deformed configuration
and the pairing correlations are accounted in the BCS formalism. 
It is known that the BCS description of the scattering of 
nucleonic pairs from bound states to the positive-energy particle 
continuum produces an unphysical component in the nucleon density
with the wrong asymptotic behavior~\cite{DFT.84,DNW.96}. 
This effect is more pronounced for neutron-rich nuclei,
for which the coupling to the particle continuum is 
particularly important. For proton-rich nuclei, however, the 
Coulomb barrier confines the protons in the interior 
of the nucleus. Therefore, the effect of the 
coupling to the continuum is weaker, and, for nuclei close 
to the proton drip line, the RMF+BCS approach can still be considered 
as a reasonable approximation providing  sufficiently
accurate solutions. Moreover, it has been shown in Ref.~\cite{NaZ.96} 
that the total energy is not affected seriously by this coupling. 
Of course, it is more desirable if pairing
correlations are described in the unified framework of the
Relativistic-Hartree-Bogoliubov (RHB) scheme 
[or Hartree-Fock-Bogoliubov (HFB) in the nonrelativistic approach], 
in which the nucleon densities 
have the correct asymptotic behavior. However, numerical codes 
for deformed RHB calculations are not yet generally available. 
Those appearing in published RHB (HFB) studies use spherical
configurations \cite{NaZ.96,VLR.98,Doba.93,Doba.95,Doba.96,LVP.98}.    
On the other hand, a detailed study of proton-rich nuclei within the 
deformed HF+BCS approach with the Skyrme effective
force SIII has been reported recently \cite{Taj.96}. 

The current paper is the first systematic study of the proton drip-line 
nuclei over a wide range of $Z$ values within the RMF+BCS model.
In Sec. II, a brief description of the RMF formalism is given,  
while in Sec. III, the results of our calculations
are presented and discussed. Ground-state properties such as  
binding energies, two-proton separation energies, 
proton root-mean-square (rms) radii, and deformation parameters that 
result from fully self-consistent RMF solutions have been calculated 
for very proton-rich nuclei near the proton drip line. Finally 
the prediction of the RMF theory for the location of the two-proton drip
line is compared with those obtained from other theoretical models. 

Strictly speaking, the proton drip line is delineated in a 
$Z$ vs. $N$ plot by nuclei with the smallest positive value of 
the proton separation energy $S_{1p}$. To derive the global drip line, 
it is necessary to perform calculations for all nuclei, especially the 
odd-$Z$ and odd-$N$ ones. The RMF calculations for these nuclei are very 
involved and take prohibitively long computing times. Therefore, this 
work deals only with even-even nuclei and with the two-proton drip line 
defined by nuclei with the smallest positive value of 
the two-proton separation energy $S_{2p}$. This restriction is not 
too severe because it can be shown that the drip lines defined by 
$S_{1p}$ and $S_{2p}$ are nearly parallel, except that nuclei specified 
by the $S_{1p}$ line tends, on the average, to have one or two fewer 
nucleons than those specified by the $S_{2p}$ line.       
%
\section{The RMF formalism}

In relativistic quantum hadrodynamics
the nucleons, described as Dirac particles, are coupled to
exchange mesons and photon through an effective Lagrangian.
The model is based on the one-boson exchange description of
the nucleon-nucleon interaction. The Lagrangian density of
the model is given by \cite{Ring.96}  

\begin{eqnarray}
{\cal L}&=&\bar\psi\left(i\gamma\cdot\partial-m\right)\psi
~+~\frac{1}{2}(\partial\sigma)^2-U(\sigma )
\nonumber\\
&&-~\frac{1}{4}\Omega_{\mu\nu}\Omega^{\mu\nu}
+\frac{1}{2}m^2_\omega\omega^2
~-~\frac{1}{4}{\vec{\rm R}}_{\mu\nu}{\vec{\rm R}}^{\mu\nu}
+\frac{1}{2}m^2_\rho\vec\rho^{\,2}
~-~\frac{1}{4}{\rm F}_{\mu\nu}{\rm F}^{\mu\nu}
\nonumber\\
&&-~g_\sigma\bar\psi\sigma\psi~-~
g_\omega\bar\psi\gamma\cdot\omega\psi~-~
g_\rho\bar\psi\gamma\cdot\vec\rho\vec\tau\psi~-~
e\bar\psi\gamma\cdot A \frac{(1-\tau_3)}{2}\psi\;.
\label{lagrangian}
\end{eqnarray}
The Dirac spinor $\psi$ denotes the nucleon with mass $m$.
The quantities 
$m_\sigma$, $m_\omega$, and $m_\rho$ are the masses of the
$\sigma$ meson, the $\omega$ meson, and the $\rho$ meson,
respectively, and $g_\sigma$, $g_\omega$, and $g_\rho$ are the
corresponding coupling constants for the mesons to the
nucleon. $U(\sigma)$ denotes the nonlinear $\sigma$
self-interaction \cite{BB.77}, 
\begin{equation}
U(\sigma)~=~\frac{1}{2}m^2_\sigma\sigma^2+\frac{1}{3}g_2\sigma^3+
\frac{1}{4}g_3\sigma^4,
\label{NL}
\end{equation}
and $\Omega^{\mu\nu}$, $\vec R^{\mu\nu}$, and $F^{\mu\nu}$
are field tensors~\cite{SW.86}.

Assuming time-reversal
symmetry and charge conservation, the coupled
equations of motion are derived from the Langrangian 
density (\ref{lagrangian}).  
The Dirac equation for the
nucleons is
\begin{equation}
\{ -i{\bf {\alpha}} \nabla + V({\bf r}) + \beta [ M + S({\bf r}) ] \}
~\psi_{i} = ~\epsilon_{i} \psi_{i} .
\end{equation}
The Klein-Gordon equations for the mesons are 
\begin{equation}
\begin{array}{ll}
\{ -\Delta + m_{\sigma}^{2} \}\sigma({\bf r})
 =& -g_{\sigma}\rho_{s}({\bf r})
-g_{2}\sigma^{2}({\bf r})-g_{3}^{3}({\bf r}),\\
\         \\
\  \{ -\Delta + m_{\omega}^{2} \} \omega_{0}({\bf r})
=& g_{\omega}\rho_{v}({\bf r}),\\
\                            \\
\  \{ -\Delta + m_{\rho}^{2} \}\rho_{0}({\bf r})
=& g_{\rho} \rho_{3}({\bf r}),\\
\                           \\
\  -\Delta A_{0}({\bf r}) = e\rho_{c}({\bf r}).
\end{array}
\end{equation}
The nucleon densities act as sources, and the contributions of
negative-energy states are neglected 
($no$-$sea$ approximation \cite{Rei.89}). 
More details on the RMF formalism can be found in 
Refs.~\cite{SW.86,Rei.89,Ser.92,Ring.96}. 

\section{ Numerical results and comments}

In this work, the Dirac equation for nucleons is solved using
the method of oscillator expansion as described in Ref. \cite{GRT.90}.
Because most of the nuclei considered here are open-shell
nuclei, both proton and neutron pairing correlations have been included.
The BCS formalism was used for the pairing with
constant pairing gaps obtained from the prescription of Ref.
\cite{MN.92}. The number of oscillator shells taken into account is 12
for fermionic and 20 for bosonic wave functions.
The effective force NL3 was adopted for the calulations
using a new version of the ``axially-deformed'' code \cite{RGL.97}. 
The parameter set NL3 has been derived recently~\cite{LKR.97} by fitting
ground-state properties of ten spherical nuclei.
Properties predicted with the NL3 effective interaction are found to
be in good agreement with experimental data \cite{LKR.97,LVR.98} for 
nuclei at and away from the line of $\beta$ stability.

The calculations have been performed for several nuclei close to the
proton drip line for the even-even isotopic chains. 
In Table I the calculated total binding energies for the three most
proton-rich isotopes close to the drip line are listed for each element
with atomic numbers  ranging from $Z=10$ to $Z=82$.
The experimental values (in parentheses), if available, are also
shown for comparison. With the exception of the binding energies for 
$^{80}$Zr, $^{100}$Sn, $^{174}$Hg, and $^{180}$Pb, which are 
from Refs. \cite{Iss.97,Cha.96,Sum.97,Uus.97,Page.96,Tot.96}, 
all other values are from the 1995 
Atomic Mass Adjustment \cite{AW.95}. The rms deviation 
between calculation and experiment is only 3.1~MeV. 
The larger differences are observed for $N \approx Z$ nuclei.     
This observation might indicate that for these nuclei additional
correlations should be taken into account \cite{Zel.96}.
In particular, proton-neutron pairing 
could have a strong influence on the masses. Proton-neutron 
short-range correlations are not included in our model.

In Fig. 1, the two-proton separation energies 
\begin{equation}
S_{2p}(Z,N) = B(Z,N) - B(Z-2,N)
\label{sep}
\end{equation}
are shown as function of the atomic number $Z$. 
In the upper panel are shown the
two-proton separation energies $S_{2p}$ for nuclei with
$Z=10-48$, while in the lower panel are shown the corresponding 
values for nuclei with $Z=48-82$. 
Each curve corresponds to a given neutron number which changes
from $N=8$ to $N=46$ (upper panel) and $N=48$ to $N=94$ (lower panel)   
in going from the left to the right of the figures.    

In Table II are listed (first column) the predictions of the RMF theory for 
the most proton rich even-even nuclei (with $10 \leq  Z \leq 82$) that are   
stable with respect to the two-proton emission, i.e.,   
$ S_{2p}(Z,N) > 0$. For comparison, the corresponding  
predictions of the HF+BCS mean-field theory (second column) with 
the effective force Skyrme SIII \cite{Taj.96} and of the 
finite-range droplet model 
(FRDM) model \cite{MNM.95,MNK.97} (third column) 
are also given. Finally in the fourth 
column the lightest experimentally known, proton-stable nuclei 
are listed for each even-$Z$ element. 
It is seen that the predictions of the various theoretical
models are in accordance in most of the cases. Whether such close agreement 
exists in the neutron-rich region is an open question.

In Table III, the predictions of the RMF theory for 
the quadrupole deformation parameter $\beta_{2}$ are shown
for all nuclei listed in the first column of Table I. 
It is seen that most of the nuclei 
close to the proton drip line are deformed, apart from those with 
magic proton (neutron) number, which are spherical or almost spherical.    
It turns out that the magic numbers maintain their character close to
the proton drip line. In Fig. 2,  
the trend of the variation of the quadropole deformation parameter 
$\beta_{2}$ of the most proton-rich even-even nuclei that are stable to  
two-proton emission is shown as a function of $Z$. 

Table III also gives the RMF predictions for the proton radii $r_{p}$. 
Unlike the other calculated ground-state properties, these $r_{p}$ values
must be treated with some caution because, near the proton 
drip line, the BCS approach may not be a sufficiently good approximation 
for estimating proton radii.       

In conclusion, a systematic study of the properties
of very proton-rich nuclei close to the drip line has been carried out. 
The location of the two-nucleon proton drip line has been predicted, 
which is in agreement with the predictions of other theoretical 
models. In 14 of 37 cases (of even-$Z$ elements), the proton drip 
line has apparently been reached in a variety of experiments. 
The existing calculations (see Table II) suggest that there are 
approximately 60 unknown isotopes of even-$Z$ elements  
in the $10 \leq  Z \leq 82$ region that are proton stable. 
The smallness of this number reflects the increased activity in this 
research area in recent years. The number of undiscovered isotopes in the 
neutron-rich side is, of course, much larger. Calculations similar to those 
reported here have been carried out by us for over 1300 even-even nuclei on 
either side of the valley of stability. These results will be reported 
separately.   

\begin{center} \bf ACKNOWLEDGMENTS \end{center} 

Helpful comments by C. Baktash, K. Rykaczewski, and W. Nazarewicz are 
acknowledged. The assistance of G. Audi in updating the experimental values 
given in Table I is also acknowledged. One of us (G.A.L) is grateful to the 
Joint Institute for Heavy-Ion Research for arranging his assignment to Oak Ridge. 
The current work was sponsored by the U.S. Department of Energy under 
Contract No. DE-AC05-96OR22464 with the  
Lockheed Martin Energy Research Corporation.


\newpage
\widetext
\begin{table}
\caption{Comparison of calculated and experimental 
binding energies (in MeV) for some very proton-rich nuclei. 
Experimental values, where available, are 
displayed in parentheses. In our notation, 
132.153~{\it 2}~$\equiv$~132.153~$\pm~0.002$, 
134.47~{\it 3}~$\equiv$~134.47~$\pm~0.03$, etc.} 
\begin{center}
\begin{tabular}{llllll}
$^{18}$Ne& 134.70 (132.153 {\it 2}) &$^{68}$Se&572.28  &$^{124}$Nd&1020.57          
               \\
$^{20}$Ne& 155.51 (160.645 {\it 1})&$^{68}$Kr&544.35         & $^{126}$Nd&1042.02         
               \\
$^{22}$Ne& 176.18 (177.770 {\it 1})&$^{70}$Kr&575.14         & $^{128}$Sm&1025.90      \\
$^{20}$Mg& 136.62 (134.47 {\it 3})&$^{72}$Kr&602.92 (607.08 {\it 28})&$^{130}$Sm&1050.02 \\
$^{22}$Mg& 166.97 (168.578 {\it 2})&$^{74}$Sr&605.02         &$^{132}$Sm&1073.29       \\
$^{24}$Mg& 194.51 (198.257 {\it 1})&$^{76}$Sr&634.86         &$^{132}$Gd&1050.66        \\
$^{22}$Si& 136.94         &$^{78}$Sr&660.08 (663.008 {\it 8})&$^{134}$Gd&1075.62        
\\
$^{24}$Si& 170.61 (172.004 {\it 20})&$^{78}$Zr&637.10         &$^{136}$Gd&1098.81         
\\
$^{26}$Si& 202.85 (206.046 {\it 3})&$^{80}$Zr&665.52 (669.9 {\it 15})&$^{136}$Dy&1075.72\\
$^{26}$S & 171.17         &$^{82}$Zr&690.59 (694.7 {\it 6})&$^{138}$Dy&1099.89        \\
$^{28}$S & 207.28 (209.41 {\it 17})&$^{82}$Mo&666.70         &$^{140}$Dy&1122.86       \\
$^{30}$S & 239.98 (243.685 {\it 4})&$^{84}$Mo&696.05         &$^{142}$Er&1123.66          
\\
$^{32}$Ar& 244.56 (246.38 {\it 5})&$^{86}$Mo&720.93 (725.8 {\it 5})&$^{144}$Er&1147.01 \\
$^{34}$Ar& 274.94 (278.721 {\it 4})&$^{86}$Ru&698.08         &$^{146}$Er&1171.18        
\\
$^{36}$Ar& 302.78 (306.716 {\it 1})&$^{88}$Ru&726.42         &$^{146}$Yb&1147.13        \\
$^{34}$Ca & 246.29        &$^{90}$Ru&755.03         &$^{148}$Yb&1172.49          
 \\
$^{36}$Ca & 280.49 (281.36 {\it 4})&$^{90}$Pd&729.27        &$^{150}$Yb&1197.32        
\\
$^{38}$Ca & 312.19 (313.122 {\it 5})&$^{92}$Pd&760.26        &$^{152}$Hf&1197.93    
    \\
$^{40}$Ti& 314.07 (314.49 {\it 16})&$^{94}$Pd&789.17         &$^{154}$Hf&1221.51          
\\
$^{42}$Ti& 347.89 (346.905 {\it 6})&$^{94}$Cd&762.49         &$^{156}$Hf&1242.72          
 \\
$^{44}$Ti& 372.30 (375.475 {\it 1})&$^{96}$Cd&794.21         & $^{156}$W&1222.58       \\
$^{44}$Cr & 350.43        &$^{98}$Cd&824.87         &$^{158}$W&1244.50        \\
$^{46}$Cr & 378.63 (381.975 {\it 20})&$^{98}$Sn&797.11        &$^{160}$W&1265.97        \\
$^{48}$Cr & 408.92 (411.462 {\it 8})&$^{100} 
$Sn&829.94 (825.2 {\it 6}) &$^{160}$Os&1244.57       \\
$^{46}$Fe& 351.34          &$^{102}$Sn&852.56       &$^{162}$Os&1267.07       \\
$^{48}$Fe& 383.65          &$^{106}$Te&874.22       &$^{164}$Os&1288.71       \\
$^{50}$Fe& 416.17 (417.70 {\it 6})&$^{108}$Te&896.94 (896.70 {\it 16})& $^{164}$Pt&1267.40     \\
$^{50}$Ni& 385.20         &$^{110}$Te&918.42 (919.44 {\it 6})&$^{166}$Pt&1289.23      \\
$^{52}$Ni& 418.66         &$^{110}$Xe&897.61         & $^{168}$Pt&1310.64     \\
$^{54}$Ni& 451.67 (453.15 {\it 5})&$^{112}$Xe&921.11 (921.67{\it 16})& $^{170}$Hg&1311.45     \\
$^{56}$Zn& 452.49         &$^{114}$Xe&943.73 &$^{172}$Hg&1333.42      \\
$^{58}$Zn& 484.68 (486.96 {\it 5})&$^{114}$Ba&921.37         &$^{174}$Hg&1353.46 (1354.74 {\it 3})\\
$^{60}$Zn& 510.89 (514.992 {\it 11})&$^{116}$Ba&946.82         &$^{176}$Pb&1354.16      \\
$^{62}$Ge& 514.11         &$^{118}$Ba&970.50 & $^{178}$Pb&1374.40     \\
$^{64}$Ge& 540.19 (545.95 {\it 26})&$^{118}$Ce&948.73         &$^{180}$Pb&1394.17 (1390.65 {\it 3})\\
$^{66}$Ge& 564.71 (569.29 {\it 4})&$^{120}$Ce&974.03         &          &             \\ 

$^{64}$Se& 514.40         &$^{122}$Ce&997.93         &          &             \\ 

$^{66}$Se& 544.10         &$^{122}$Nd&975.49         &          &             \\
\end{tabular}
\end{center}
\label{TabA}
\end{table}

\newpage
\narrowtext
\begin{table}
\caption {Predictions of the RMF theory for the most proton-rich, even-even, 
proton-stable nuclei with $10 \leq  Z \leq 82$. 
Predictions of the HF+BCS mean-field theory 
and of the FRDM model are also shown. In the last column
are listed the most proton-rich nuclei known experimentally.} 
\begin{center}
\begin{tabular}{cccc}

Calculation & Calculation & Calculation &  \\
RMF+BCS  & HF+BCS \protect{\cite{Taj.96}}  
& FRDM \protect{\cite{MNM.95,MNK.97}} & Experiment \\
(NL3) & (SIII) & & \\
\hline
$^{18}$Ne &$^{18}$Ne      &  $^{18}$Ne   & $^{16}$Ne \\
$^{20}$Mg &$^{20}$Mg      &  $^{20}$Mg   & $^{20}$Mg \\
$^{22}$Si &$^{22}$Si      &  $^{24}$Si   & $^{22}$Si \\
$^{26}$S &$^{26}$S      &  $^{28}$S   & $^{27}$S \\
$^{32}$Ar &$^{32}$Ar      &  $^{32}$Ar   & $^{31}$Ar \\
$^{34}$Ca &$^{34}$Ca      &  $^{36}$Ca   & $^{35}$Ca \\
$^{40}$Ti &$^{40}$Ti      &  $^{40}$Ti   & $^{39}$Ti \\
$^{44}$Cr &$^{43}$Cr      &  $^{44}$Cr   & $^{43}$Cr \\
$^{46}$Fe &$^{46}$Fe      &  $^{48}$Fe   & $^{45}$Fe \\
$^{50}$Ni &$^{50}$Ni      &  $^{50}$Ni   & $^{49}$Ni \\
$^{56}$Zn &$^{56}$Zn      &  $^{56}$Zn   & $^{57}$Zn \\
$^{62}$Ge &$^{60}$Ge      &  $^{62}$Ge   & $^{61}$Ge \\
$^{64}$Se &$^{64}$Se      &  $^{66}$Se   & $^{66}$Se \\
$^{68}$Kr &$^{68}$Kr      &  $^{70}$Kr   & $^{71}$Kr \\
$^{74}$Sr &$^{72}$Sr      &  $^{74}$Sr   & $^{73}$Sr \\
$^{78}$Zr &$^{76}$Zr      &  $^{78}$Zr   & $^{79}$Zr \\
$^{82}$Mo &$^{80}$Mo      &  $^{84}$Mo   & $^{83}$Mo \\
$^{86}$Ru &$^{82}$Ru      &  $^{86}$Ru   & $^{87}$Ru \\
$^{90}$Pd &$^{88}$Pd      &  $^{90}$Pd   & $^{91}$Pd \\
$^{94}$Cd &$^{92}$Cd      &  $^{94}$Cd   & $^{97}$Cd \\
$^{98}$Sn &$^{96}$Sn      &  $^{98}$Sn   & $^{100}$Sn \\
$^{106}$Te &$^{108}$Te      &  $^{108}$Te   & $^{106}$Te \\
$^{110}$Xe &$^{110}$Xe      &  $^{110}$Xe   & $^{110}$Xe \\
$^{114}$Ba &$^{114}$Ba      &  $^{114}$Ba   & $^{114}$Ba \\
$^{118}$Ce &$^{118}$Ce      &  $^{118}$Ce   & $^{121}$Ce \\
$^{122}$Nd &$^{122}$Nd      &  $^{122}$Nd   & $^{127}$Nd \\
$^{128}$Sm &$^{128}$Sm      &  $^{128}$Sm   & $^{131}$Sm \\
$^{132}$Gd &$^{132}$Gd      &  $^{134}$Gd   & $^{135}$Gd \\
$^{136}$Dy &$^{136}$Dy      &  $^{138}$Dy   & $^{141}$Dy \\
$^{142}$Er &$^{142}$Er      &  $^{144}$Er   & $^{145}$Er \\
$^{146}$Yb &$^{148}$Yb      &  $^{148}$Yb   & $^{150}$Yb \\
$^{152}$Hf &$^{152}$Hf      &  $^{154}$Hf   & $^{154}$Hf \\
$^{156}$W  &$^{156}$W      &  $^{158}$W   & $^{158}$W \\
$^{160}$Os &$^{162}$Os      &  $^{162}$Os   & $^{162}$Os \\
$^{164}$Pt &$^{166}$Pt    &  $^{170}$Pt   & $^{166}$Pt \\
$^{170}$Hg &$^{172}$Hg    &  $^{174}$Hg   & $^{174}$Hg \\
$^{176}$Pb &$^{176}$Pb    &  $^{180}$Pb   & $^{180}$Pb \\
\end{tabular}
\end{center}
\label{TabB}
\end{table}

\newpage
\widetext
\begin{table}
\caption{Predictions of the RMF theory for the proton radii ({\it r}$_{p}$) 
and quadrupole deformation parameters ($\beta_{2}$) for proton-rich nuclei close
to the proton drip line.}
\begin{center}
\begin{tabular}{ccdccdccd}
Nucleus&{\it r}$_{p}$&$\beta_{2}$&Nucleus&{\it r}$_{p}$&$\beta_{2}$&Nucleus&{\it r}$_{p}$&$\beta_{2}$\\
\hline
$^{18}$Ne&2.959&0.001&$^{68}$Se&4.010&--0.285&$^{124}$Nd&4.854&0.341\\
                 
$^{20}$Ne&2.911&0.186&$^{68}$Kr&4.075&--0.274&$^{126}$Nd&4.862&0.339\\
                 
$^{22}$Ne&2.892&0.350&$^{70}$Kr&4.087&--0.310&$^{128}$Sm&4.905&0.346\\

$^{20}$Mg&3.120&0.002&$^{72}$Kr&4.103&--0.358&$^{130}$Sm&4.911&0.343\\

$^{22}$Mg&3.076&0.356&$^{74}$Sr&4.195& 0.387&$^{132}$Sm&4.920&0.341\\

$^{24}$Mg&3.021&0.416&$^{76}$Sr&4.207& 0.410&$^{132}$Gd&4.954&0.346\\

$^{22}$Si&3.266&--0.001&$^{78}$Sr&4.213& 0.417&$^{134}$Gd&4.959&0.344\\

$^{24}$Si&3.186& 0.230&$^{78}$Zr&4.272& 0.422&$^{136}$Gd&4.985&0.359\\

$^{26}$Si&3.133& 0.320&$^{80}$Zr&4.276& 0.437&$^{136}$Dy&4.998&0.345\\

$^{26}$S &3.332& 0.001&$^{82}$Zr&4.205&--0.232&$^{138}$Dy&5.012&0.346\\

$^{28}$S &3.270& 0.268&$^{82}$Mo&4.256&--0.230&$^{140}$Dy&5.017&0.326\\

$^{30}$S &3.205&--0.224&$^{84}$Mo&4.258&--0.247&$^{142}$Er&5.036&0.297\\

$^{32}$Ar&3.333&--0.145&$^{86}$Mo&4.241& 0.003&$^{144}$Er&5.033&0.257\\

$^{34}$Ar&3.316&--0.176&$^{86}$Ru&4.308&--0.244&$^{146}$Er&5.014&--0.207\\

$^{36}$Ar&3.318&--0.207&$^{88}$Ru&4.296& 0.107&$^{146}$Yb&5.051&--0.251\\

$^{34}$Ca&3.393& 0.000&$^{90}$Ru&4.294& 0.113&$^{148}$Yb&5.048&--0.207\\
   
$^{36}$Ca&3.375& 0.000&$^{90}$Pd&4.339& 0.109&$^{150}$Yb&5.049&--0.180\\

$^{38}$Ca&3.373& 0.000&$^{92}$Pd&4.336& 0.112&$^{152}$Hf&5.078&--0.163\\
    
$^{40}$Ti&3.524& 0.001&$^{94}$Pd&4.330& 0.071&$^{154}$Hf&5.062&--0.009\\
  
$^{42}$Ti&3.506& 0.000&$^{94}$Cd&4.371& 0.071&$^{156}$Hf&5.089&--0.090\\
   
$^{44}$Ti&3.497& 0.000&$^{96}$Cd&4.363& 0.003&$^{156}$W& 5.094&--0.006\\

$^{44}$Cr&3.607& 0.000&$^{98}$Cd&4.357& 0.001&$^{158}$W& 5.117&--0.066\\
        
$^{46}$Cr&3.586&--0.004&$^{98}$Sn&4.394& 0.001&$^{160}$W& 5.143& 0.110\\

$^{48}$Cr&3.603& 0.225&$^{100}$Sn&4.388&0.001&$^{160}$Os&5.142& 0.022\\

$^{46}$Fe&3.666& 0.003&$^{102}$Sn&4.411&0.002&$^{162}$Os&5.166&--0.083\\

$^{48}$Fe&3.649& 0.084&$^{106}$Te&4.514&0.120&$^{164}$Os&5.189& 0.106\\

$^{50}$Fe&3.655& 0.212&$^{108}$Te&4.535&0.142&$^{164}$Pt&5.193&--0.056\\

$^{50}$Ni&3.673& 0.000&$^{110}$Te&4.553&0.153&$^{166}$Pt&5.212& 0.061\\

$^{52}$Ni&3.654& 0.001&$^{110}$Xe&4.600&0.177&$^{168}$Pt&5.229&0.066\\

$^{54}$Ni&3.639& 0.000&$^{112}$Xe&4.617&0.195&$^{170}$Hg&5.254&--0.006\\

$^{56}$Zn&3.810& 0.154&$^{114}$Xe&4.636&0.221&$^{172}$Hg&5.270&--0.001\\

$^{58}$Zn&3.769&--0.001&$^{114}$Ba&4.680&0.230&$^{174}$Hg&5.283&--0.030\\

$^{60}$Zn&3.800& 0.170&$^{116}$Ba&4.717&0.285&$^{176}$Pb&5.303& 0.000\\

$^{62}$Ge&3.888& 0.197&$^{118}$Ba&4.731&0.295&$^{178}$Pb&5.313& 0.001\\

$^{64}$Ge&3.904& 0.217&$^{118}$Ce&4.783&0.315&$^{180}$Pb&5.322& 0.003\\

$^{66}$Ge&3.931&--0.261&$^{120}$Ce&4.796&0.326&          &      &     \\

$^{64}$Se&3.976& 0.205&$^{122}$Ce&4.805&0.328&          &      &      \\

$^{66}$Se&3.997&--0.265&$^{122}$Nd&4.847&0.341&          &      &       \\
\end{tabular}
\end{center}
\label{TabC}
\end{table}

\begin{figure}
\caption{Calculated two-proton separation energies $S_{2p}$ for the $N=8-94$ isotones 
as a function of the proton number {\it Z}.} 

\label{figA}
\end{figure}

\begin{figure}
\caption{Calculated quadrupole deformation parameters $\beta_{2}$ of the most proton-rich, 
proton-stable, even-even nuclei with proton numbers from $Z=10$ to $Z=82$.} 
\label{figB}
\end{figure}



\begin{references}

\bibitem{Rkl.92} E. Roeckl, Rep. Prog. Phys. {\bf 55}, 1661 (1992).

\bibitem{MS.93} A. Mueller and B. Sherril, Annu. Rev. Nucl. Part. Sci.
                 {\bf 43}, 529 (1993).

\bibitem{Gei.95} H. Geissel, G. M\"unzenberg, and K. Riisager, 
Annu. Rev. Nucl. Part. Sci. {\bf 45}, 163 (1995).  

\bibitem{Han.95} P. G. Hansen, A. S. Jensen, and B. Jonson, 
Annu. Rev. Nucl. Part. Sci. {\bf 45}, 591 (1995). 


\bibitem{Tan.95} I. Tanihata, Progr. Part. Nucl. Phys. {\bf 35}, 505 
                  (1996). 

\bibitem{Ver.96} J. Vervier, Prog. Part. Nucl. Phys. {\bf 37},
               435 (1996).

\bibitem{Proc.96} {\it Proceedings of the Fourth International 
Conference on Radioactive Nuclear Beams (Omiya, Japan)}, 
edited by S. Kubono, T. Kobayashi, and I. Tanihata, 
Nucl. Phys. {\bf A616} (1997).

\bibitem{Orm.96} W. E. Ormand, Phys. Rev. C {\bf 53}, 214 (1996).

\bibitem{Orm.97} W. E. Ormand, Phys. Rev. C {\bf 55}, 2407 (1997).

\bibitem{NaZ.96} W. Nazarewicz, J. Dobaczewski, 
                T. R. Werner, J. A. Maruhn,
                P.-G. Reinhard, K. Rutz, C. R. Chinn, A. S. Umar, and 
                M. R. Strayer, Phys. Rev. C {\bf 53}, 740 (1996).

\bibitem{VLR.98} D. Vretenar, G.A. Lalazissis, and P. Ring, 
                Phys. Rev. C, in print.

\bibitem{Det.90} C. D\' etraz, R. Anne, P. Bricault, 
D. Guillemaud-Mueller, M. Lewitowicz, A. C. Mueller, Yu Hu Zhang, 
V. Borrel, J. C. Jacmart, F. Pougheon, A. Richard, D. Bazin, J. P. Dufour, 
A. Fleury, F. Hubert, and M. S. Pravikoff,  
                Nucl. Phys. {\bf A519}, 529 (1990).

\bibitem{Bor.92} V. Borrel, R. Anne, D. Bazin, C. Borcea, G. G. Chubarian, 
R. Del Moral, C. D\'etraz, S. Dogny, J. P. Dufour, L. Faux, A. Fleury, 
L. K. Fifield, D. Guillemaud-Mueller, F. Hubert, E. Kashy, M. Lewitowicz, 
C. Marchand, A. C. Mueller, F. Pougheon, M. S. Pravikoff, M. G. Saint-Laurent, 
and O. Sorlin, Z. Phys. A {\bf 344}, 135 (1992).

\bibitem{Bla.94} B. Blank, S. Andriamonje, R. Del Moral, 
J. P. Dufour, A. Fleury, T. Josso, M. S. Pravikoff, 
S. Czajkowski, Z. Janas, A. Piechaczek, E. Roeckl, K.-H. Schmidt,   
K. S\"ummerer, W. Trinder, M. Weber, T. Brohm, A. Grewe, E. Hanelt, 
A. Heinz, A. Junghans, C. R\"ohl, S. Steinh\"auser, B. Voss, 
and M. Pf\"utzner, Phys. Rev. C {\bf 50}, 2398 (1994).

\bibitem{Bla.96} B. Blank, S. Czajkowski, F. Davi, R. Del Moral, 
J. P. Dufour, A. Fleury, C. Marchand, M. S. Pravikoff, J. Benlliure, 
F. Bo\'ue, R. Collatz, A. Heinz, M. Hellstr\"om, Z. Hu, E. Roeckl, 
M. Shibata, K. S\"ummerer, Z. Janas, M. Karny, M. Pf\"utzner, and 
M. Lewitowicz, Phys. Rev. Lett. {\bf 77}, 2893 (1996).

\bibitem{Woods.97} P. J. Woods and C. N. Davids, Annu. Rev. Nucl. Part. Sci.
                 {\bf 47}, 541 (1997).

\bibitem{SW.86} B. D. Serot and J. D. Walecka,
    Adv. Nucl. Phys. {\bf 16}, 1 (1997). 

\bibitem{Rei.89} P. G. Reinhard, Rep. Prog. Phys. {\bf 52}, 439 (1989).

\bibitem{Ser.92} B. D. Serot, Rep. Prog. Phys. {\bf 55}, 1855 (1992).

\bibitem{Ring.96} P. Ring, Prog. Part. Nucl. Phys. {\bf 37}, 193 (1996).

\bibitem{DFT.84} J. Dobaczewski, H. Flocard, and J. Treiner,
        Nucl. Phys. {\bf A422}, 103 (1984).

\bibitem{DNW.96} J. Dobaczewski, W. Nazarewicz, T. R. Werner,
        J.-F. Berger, C. R. Chinn, and J. Decharg\' e,
        Phys. Rev. C {\bf 53}, 2809 (1996).

\bibitem{Doba.93} R. Smola\' nczuk and  J. Dobaczewski,
                  Phys. Rev. C {\bf 48}, R2166 (1993).

\bibitem{Doba.95} W. Nazarewicz, J. Dobaczewski, and 
                T. R. Werner,  Phys. Scr. {\bf T56}, 9 (1995).

\bibitem{Doba.96} J. Dobaczewski,  W. Nazarewicz, and  
                T. R. Werner,  Z. Phys. A {\bf 354}, 27 (1996).

\bibitem{LVP.98} G. A. Lalazissis, D. Vretenar, W. P\"oschl,
        and P. Ring, Nucl. Phys. A, in print. 

\bibitem{Taj.96}N. Tajima, S. Takahara, and N. Onishi, Nucl. Phys.
       {\bf A603}, 23 (1996).

\bibitem{BB.77} J. Boguta and A. R. Bodmer, Nucl. Phys. {\bf A292}, 413 (1977).

\bibitem{GRT.90} Y. K. Gambhir, P. Ring, and A. Thimet, Ann. Phys. (N.Y.) 
{\bf 198}, 132 (1990).

\bibitem{MN.92} P. M\"oller and J. R. Nix,  Nucl. Phys. {\bf A536}, 20 (1992).

\bibitem{RGL.97}  P. Ring, Y. K. Gambhir, and G. A. Lalazissis,
Comput. Phys. Commun. {\bf 105}, 77 (1997).

\bibitem{LKR.97} G. A. Lalazissis, J. K\"onig, and P. Ring, 
        Phys. Rev. C {\bf 55}, 540 (1997).

\bibitem{LVR.98} G. A. Lalazissis, D. Vretenar, and P. Ring, 
                Phys. Rev. C, in print. 

\bibitem{Iss.97} S. Issmer, M. Fruneau, J. A. Pinston, M. Asghar, D. Barn\'eoud, 
J. Genevey, Th. Kerscher, and K. E. G. L\"obner, The European Physical Journal, in print.

\bibitem{Cha.96} M. Chartier, G. Auger, W. Mittig, A. L\'{e}pine-Szily, 
L. K. Fifield, J. M. Casandjian, M. Chabert, J. Ferm\'{e}, A. Gillibert, 
M. Lewitowicz, M. Mac Cormick, M. H. Moscatello, O. H. Odland, N. A. Orr, 
G. Politi, C. Spitaels, and A. C. C. Villari, Phys. Rev. Lett. {\bf 77}, 
2400 (1996). 

\bibitem{Sum.97} K. S\"ummerer, R. Schneider, T. Faestermann, J. Friese, 
H. Geissel, R. Gernh\"auser, H. Gilg, F. Heine, J. Homolka, P. Kienle, 
H. J. K\"orner, G. M\"unzenberg, J. Reinhold, and K. Zeitelhack, 
Nucl. Phys. {\bf A616}, 341c (1997).
 
\bibitem{Uus.97} J. Uusitalo, M. Leino, R. G. Allatt, T. Enqvist, 
K. Eskola, P. T. Greenlees, 
S. Hurskanen, A. Keenan, H. Kettunen, P. Kuusiniemi, R. D. Page, 
and W. H. Trzaska, Z. Phys. A {\bf 358}, 375 (1997).
 
\bibitem{Page.96} R. D. Page, P. J. Woods, R. A. Cunningham, T. Davinson, 
N. J. Davis, A. N. James, K. Livingston, P. J. Sellin, and A. C. Shotter, 
Phys. Rev. C {\bf 53}, 660 (1996).
 
\bibitem{Tot.96} K. S. Toth, J. C. Batchelder, D. M. Moltz, and 
J. D. Robertson, Z. Phys. A {\bf 355}, 225 (1996). 

\bibitem{AW.95} G. Audi and A. H. Wapstra, Nucl. Phys. {\bf A595}, 409 (1995).

\bibitem{Zel.96} N. Zeldes, in {\it Handbook of Nuclear Properties}, 
edited by D. Poenaru and W. Greiner (Clarendon, Oxford, 1996), p. 13. 

\bibitem{MNM.95} P. M\"oller, J. R. Nix, W. D. Myers, and W. J. Swiatecki,
    At. Data Nucl. Data Tables {\bf 59}, 185 (1995).

\bibitem{MNK.97} P. M\"oller, J. R. Nix, and K.-L. Kratz,
   At. Data Nucl. Data Tables {\bf 66}, 131 (1997).


\end{references}
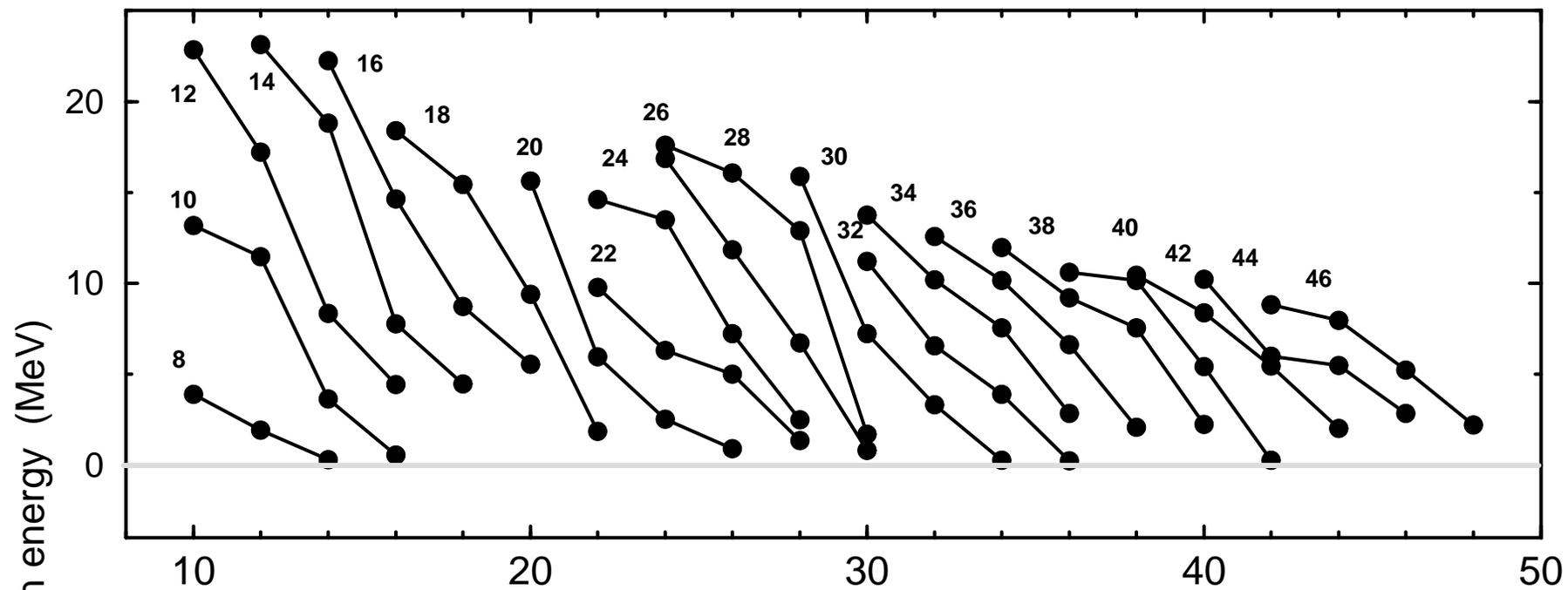
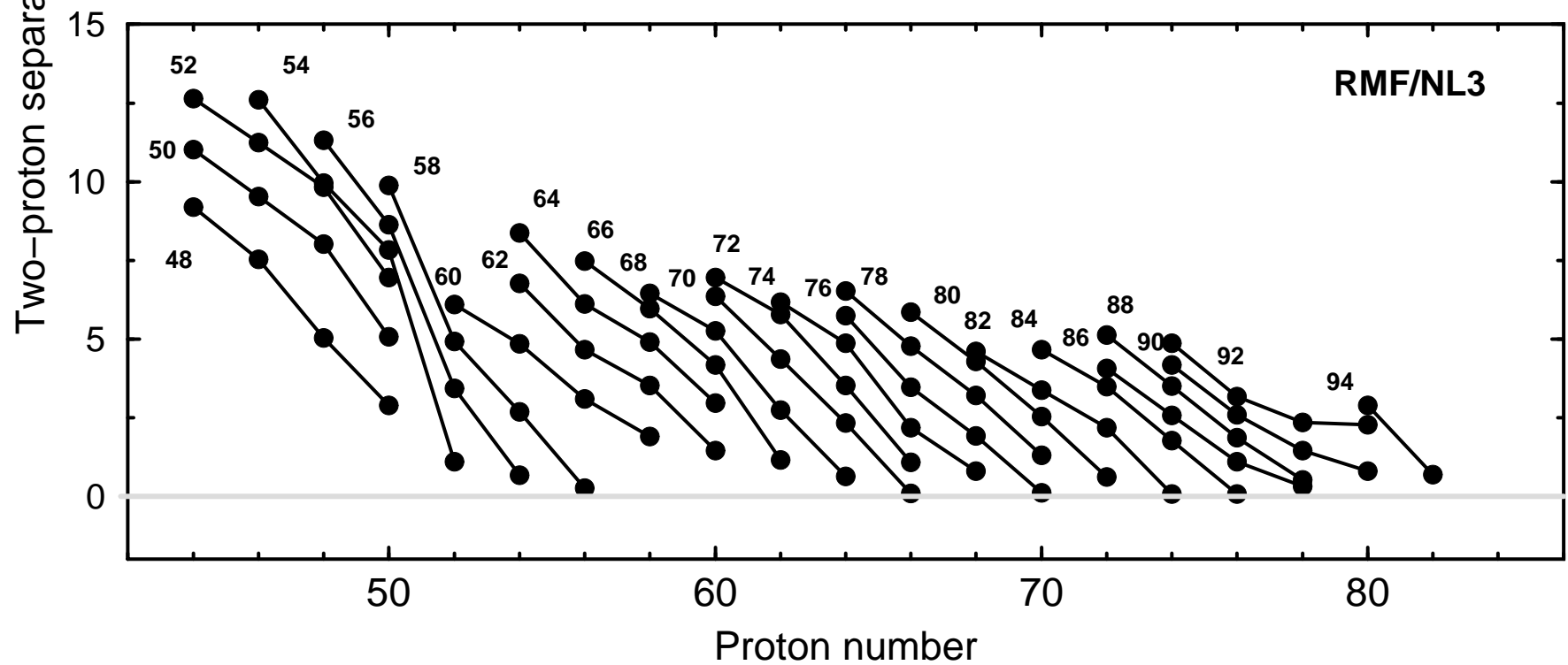
\end{document}